\newcounter{myctr}
\def\myitem{\refstepcounter{myctr}\bibfont\noindent\ifnum\themyctr>9\else\phantom{0}\fi\hangindent17pt\themyctr.\enskip}

\documentclass{ws-ijqi}
\usepackage{dcolumn}
\usepackage{bm}
\newcommand{\ket}[1]{|#1\rangle}
\newcommand{\bra}[1]{\langle#1|}

\newcommand{\scal}[2]{\langle#1|#2\rangle}

\begin{document}

\markboth{R. Lo Franco, G. Compagno, A. Messina, A. Napoli}
{Quantum computation with generalized binomial states in cavity QED}

\catchline{}{}{}{}{}

\title{QUANTUM COMPUTATION WITH GENERALIZED BINOMIAL STATES IN CAVITY QUANTUM ELECTRODYNAMICS}

\author{Rosario Lo Franco, Giuseppe Compagno, Antonino Messina, and Anna Napoli}

\address{CNISM and Dipartimento di Scienze Fisiche ed Astronomiche,
Universit\`{a} di Palermo,\\ via Archirafi 36, 90123 Palermo, Italy\\
lofranco@fisica.unipa.it}

\maketitle

\begin{history}
\received{Day Month Year}
\revised{Day Month Year}
\end{history}

\begin{abstract}
We study universal quantum computation in the cavity quantum electrodynamics (CQED) framework exploiting two orthonormal two-photon generalized binomial states as qubit and dispersive interactions of Rydberg atoms with high-$Q$ cavities. We show that an arbitrary qubit state may be generated and that controlled-NOT and 1-qubit rotation gates can be realized via standard atom-cavity interactions.
\end{abstract}


\section{Introduction}
Cavity quantum electrodynamics (CQED) has been shown to be suitable to quantum computation processing\cite{nielsenchuang} thanks to the high quality factors $Q$ of cavities, control of atom-cavity interactions and long lifetimes of Rydberg atoms\cite{harochebook}. In this context, the information can be stored and processed by atoms and photons representing the quantum bits (qubits)\cite{maitre1997PRL} and two approaches can be distinct: the ``microwave way'', where photons are confined in cavities and atoms are used to carry out the information between the cavities; the ``optical way'', where atoms are very slow or even trapped within the cavities and information is carried out by photons\cite{grangier2000FortPhys}. Following the first approach, a CQED scheme to obtain two-bit universal quantum logic gates has been proposed\cite{domokos1995PRA} and a quantum phase gate realized\cite{rausch1999PRL}, while a controlled-NOT (CNOT) gate was constructed by the optical approach\cite{barenco1995PRL,monroe1995PRL}. These schemes use resonant interactions between two-level Rydberg atoms and cavities having zero or one photon only.

Recently, interest has arisen to the coherent states and their usefulness to realize universal quantum computation in the quantum optics context\cite{jeong2002PRA,gilchrist2004JPB,jeong2007}. There, the qubit is represented by two coherent states $\ket{\alpha}$ and $\ket{-\alpha}$ of $\pi$-phase difference and the quantum logic operations, based on quantum teleportation, are realized by beam-splitters, non-linear crystals and phase-shifters\cite{jeong2002PRA}. Apart the difficulty of performing a teleportation protocol, a drawback of using coherent states is their intrinsic non orthogonality, with the consequent request of a large photon number. This problem can be overcome by moving to the microwave way of CQED and using $N$-photon generalized binomial states of electromagnetic field ($N$GBSs)\cite{stoler1985OptActa,vidiella1994PRA} stored in high-$Q$ cavities. In fact these states, that interpolate between the coherent and the number state, have the feature that to each state it corresponds another exactly orthogonal\cite{lofranco2005PRA} for any value of the maximum photon number $N$. Moreover, they can be efficiently generated by standard resonant atom-cavity interactions\cite{lofranco2006PRA,lofranco2007PRA}.

In this paper we show that universal quantum computation, that is a set of CNOT and 1-qubit rotation gates\cite{nielsenchuang}, can be realized without any teleportation protocol by exploiting two orthogonal $2$GBSs as qubit stored in cavities and dispersively interacting with Rydberg atoms.

\section{\label{GBS}Generalized binomial states}
The normalized $N$-photon generalized binomial state ($N$GBS) is defined as\cite{stoler1985OptActa}
\begin{equation}\label{NGBS}
\ket{N,p,\phi}=\sum_{n=0}^N\left[{N\choose n}p^{n}(1-p)^{N-n}\right]^{1/2}e^{in\phi}\ket{n},
\end{equation}
where $0\leq p\leq1$ is the probability of single photon occurrence and $\phi$ the mean phase\cite{vidiella1994PRA}. The $N$GBS of Eq.~(\ref{NGBS}) is equal to the vacuum state $\ket{0}$ when $p=0$ and to the number state $\ket{N}$ when $p=1$. For $N\rightarrow\infty$ and $p\rightarrow0$, so that $Np=\textrm{cost}\equiv|\alpha|^2$, the $N$GBS becomes the Glauber coherent state $\ket{|\alpha|e^{i\phi}}$. In this sense, a $N$GBS interpolates between the number and the coherent state.

Moreover, whatever $\ket{N,p,\phi}$ defined by Eq.~(\ref{NGBS}) is, the state $\ket{N,1-p,\phi+\pi}$ is such that the orthogonality property $\scal{N,p,\phi}{N,1-p,\phi+\pi}=0$ holds\cite{lofranco2005PRA}.

\section{\label{Hmodellogicalqubit}Atom-cavity dispersive interaction and the logical qubit}
\subsection{Hamiltonian model}
\begin{figure}[pb]
\centerline{\psfig{file=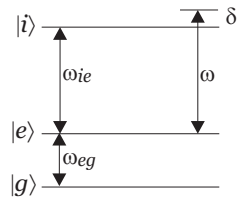}}
\vspace*{8pt}
\caption{\label{fig:levels}Atomic levels configuration. $\delta=\omega-\omega_{ie}$ is the detuning between the transition of the two upper atomic levels $\ket{i}$, $\ket{e}$ and the cavity mode frequency.}
\end{figure}

To realize our universal quantum gates we consider a Rydberg atom that crosses a cavity dispersively interacting with the field state stored inside it. The atom can be thought as an effective three-level atom, with levels $\ket{g}$, $\ket{e}$ and $\ket{i}$ as illustrated in Fig.~\ref{fig:levels}, and the cavity has a mode frequency $\omega$ slightly different from that of the transition $\ket{e}\rightarrow\ket{i}$, $\omega_{ie}$, of an amount $\delta=\omega-\omega_{ie}$. The level $\ket{g}$, the ground state of the atom, is unaffected by the atom-cavity coupling\cite{harochebook}.
It is known that, if $|\delta|\gg\Omega$ where $\Omega$ is the Rabi frequency between the cavity mode and the transition $\ket{e}\rightarrow\ket{i}$, the effective atom-cavity coupling is described by the interaction Hamiltonian\cite{harochebook} $H_\textrm{I}=(\hbar\Omega^2/\delta)a^\dag a\sigma_{eg}^+\sigma_{eg}^-$, where $a,a^\dag$ are the photon annihilation and creation operators while $\sigma_{eg}^-=\ket{g}\bra{e}$, $\sigma_{eg}^+=\ket{e}\bra{g}$. The effect of this dispersive atom-cavity interaction on a generic cavity state $\ket{\psi}=\sum c_n\ket{n}$ is obtained, in interaction picture, by applying the operator $e^{-iH_\textrm{I}t/\hbar}$ to the total atom-cavity state and gives the evolutions
\begin{equation}\label{effectofH}
\ket{g}\ket{\psi}\stackrel{H_\textrm{I}}{\rightarrow}\ket{g}\ket{\psi},\quad
\ket{e}\ket{\psi}\stackrel{H_\textrm{I}}{\rightarrow}\ket{e}\sum c_n e^{-in\Omega^2t/\delta}\ket{n}.
\end{equation}

Let us take the cavity state initially prepared in the $N$GBS $\ket{N,1/2,\phi}$ and the atom-cavity interaction time $t$ such that $\Omega^2t/\delta=\pi$. Exploiting Eq.~(\ref{effectofH}) and taking into account Eq.~(\ref{NGBS}) as well as the fact that $\ket{N,1/2,\phi-\pi}=\ket{N,1/2,\phi+\pi}$, we immediately obtain
\begin{equation}\label{effectofHonGBS}
\ket{g}\ket{N,1/2,\phi}\stackrel{H_\textrm{I}}{\rightarrow}\ket{g}\ket{N,1/2,\phi},\quad
\ket{e}\ket{N,1/2,\phi}\stackrel{H_\textrm{I}}{\rightarrow}\ket{e}\ket{N,1/2,\phi+\pi}.
\end{equation}
Thus, the initial $N$GBS remains unchanged if the atom is in the ground state $\ket{g}$, while it is transformed to its orthogonal $\ket{N,1/2,\phi+\pi}$ if the atom is in the excited state $\ket{e}$. We shall refer to the dynamics due to the dispersive interaction (DI) of Eq.~(\ref{effectofHonGBS}) as the $\pi$-DI.

\subsection{\label{thelogicalqubit}The logical qubit preparation}
Our protocols for the quantum gates will require the preparation and also the measurement of $N$GBSs inside the cavity and it is known that for $N=1,2$ this is achievable efficiently by standard resonant atom-cavity interactions\cite{lofranco2005PRA,lofranco2006PRA}. Hereafter, we consider two orthogonal 2GBSs and identify them as basis states of a logical qubit $\ket{0_L},\ket{1_L}$, that is $\ket{2,1/2,\phi}\equiv\ket{\phi}=\ket{0_L}$, $\ket{2,1/2,\phi+\pi}\equiv\ket{\phi+\pi}=\ket{1_L}$.

An arbitrary qubit state $\ket{\psi}=a\ket{\phi}+b\ket{\phi+\pi}$ can be prepared by the $\pi$-DI above, the application of two opportune Ramsey zones before and after the cavity and the final measurement of the atomic state. We recall that a Ramsey zone provides a resonant interaction of an atom with a classical field (laser) that in turn allows the following transformations:
\begin{eqnarray}\label{ramseyeq}
\ket{g}\stackrel{R_{\theta,\varphi}}{\rightarrow}\cos(\theta/2)\ket{g}+e^{-i\varphi}\sin(\theta/2)\ket{e},\
\ket{e}\stackrel{R_{\theta,\varphi}}{\rightarrow}\cos(\theta/2)\ket{e}-e^{i\varphi}\sin(\theta/2)\ket{g},
\end{eqnarray}
where $R_{\theta,\varphi}$ indicates the dependence on the parameters $\theta$ (``Ramsey pulse'') and $\varphi$, which are fixed by adjusting the classical field amplitude and the atom-field interaction time. The scheme to obtain the qubit state $\ket{\psi}=a\ket{\phi}+b\ket{\phi+\pi}$ requires the following steps: (i) the initial preparation, by a first Ramsey zone, of the atomic state $\ket{\chi}=a\ket{g}+b\ket{e}$ and of the cavity in the logic state $\ket{\phi}$; (ii) the $\pi$-DI between atom and cavity; (iii) the application of a second Ramsey zone $R_{\pi/2,0}$ after the atom has come out of the cavity; (iv) the final measurement of the atomic state. If the outcome is $\ket{e}$, occurring with a probability of $50\%$, the procedure ends successfully. If the measurement outcome is $\ket{g}$, the qubit state obtained is instead $a\ket{\phi}-b\ket{\phi+\pi}$. A deterministic procedure is also applicable to obtain the qubit state $\ket{\psi}$ that exploits the resonant interaction of two consecutive two-level atoms with the cavity initially in the vacuum state\cite{lofranco2007PRA}.

\section{\label{CNOTsection}Controlled-NOT gate scheme}
We now show how a CNOT gate can be realized by using 2GBSs and $\pi$-DI. A CNOT operation requires two qubits, namely the control and target qubits. If the control qubit is $\ket{0}$, the target qubit is unchanged, while it is flipped if the control is $\ket{1}$. The final target state can be thus written $\ket{c\oplus t}$. Since the quantum gate is coherent, it also acts on two general qubit states $\ket{\psi_t}=a\ket{0_t}+b\ket{1_t}$ (target) and $\ket{\chi_c}=c\ket{0_c}+d\ket{1_c}$ (control) giving
\begin{equation}\label{cnotoperation}
\ket{\chi_c}\ket{\psi_t}\stackrel{\textrm{CNOT}}{\longrightarrow}ac\ket{0_c0_t}+bc\ket{0_c1_t}+ad\ket{1_c1_t}+bd\ket{1_c0_t}.
\end{equation}

\begin{figure}[pb]
\centerline{\psfig{file=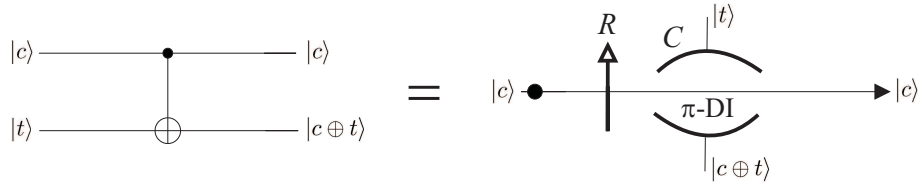}}
\vspace*{8pt}
\caption{\label{fig:CNOT}Scheme for realizing a controlled-NOT gate with 2GBSs in CQED. $\ket{c}=\{\ket{g},\ket{e}\}$ is the control qubit represented by the two levels of a Rydberg atom, while $\ket{t}=\{\ket{\phi}=\ket{0_L},\ket{\phi+\pi}=\ket{1_L}\}$ is the target qubit represented by the two orthogonal 2GBSs. $R$ is the Ramsey zone and $\pi$-DI indicates the relevant atom-cavity dispersive interaction.}
\end{figure}

In our case, this result can be realized by the $\pi$-DI of Eq.~(\ref{effectofHonGBS}) taking the two orthogonal 2GBSs $\ket{\phi},\ket{\phi+\pi}$ as target qubit and the atom as control qubit, with $\ket{g}=\ket{0_c}$ and $\ket{e}=\ket{1_c}$. The scheme is sketched in Fig.~\ref{fig:CNOT}. In fact, it is readily seen that
\begin{equation}\label{cnot2GBS}
(c\ket{g}+d\ket{e})(a\ket{\phi}+b\ket{\phi+\pi})\stackrel{\pi\textrm{-DI}}{\longrightarrow}
ac\ket{g,\phi}+bc\ket{g,\phi+\pi}+ad\ket{e,\phi+\pi}+bd\ket{e,\phi},
\end{equation}
and this transformation just coincides with the CNOT gate operation of Eq.~(\ref{cnotoperation}). Thus, a CNOT gate can be realized in a very simple way in the CQED framework by exploiting 2GBSs and an opportune dispersive interaction of a control Rydberg atom with the cavity field.

It is known that for a universal gate operation a 1-qubit rotation is required besides the CNOT gate\cite{monroe1995PRL}. In the following section we will see how this 1-qubit rotation can be achieved.

\section{\label{singlequbitrotation}1-qubit rotation scheme}
In this section we describe the CQED scheme to realize a 1-qubit rotation gate by 2GBSs and $\pi$-DI.

The relevant rotations of the logical qubit state $\ket{\psi}=a\ket{\phi}+b\ket{\phi+\pi}$ are those about the axis $\mathbf{u}=(-\sin\varphi,\cos\varphi,0)$ and about the axis $z$ represented respectively, in the computational basis $\{\ket{\phi}=\ket{0_L},\ket{\phi+\pi}=\ket{1_L}\}$, by the matrices
\begin{equation}\label{rotationmatrices}
U_\mathbf{u}(\theta/2)=\left(\begin{array}{cc}\cos(\theta/2)& -e^{i\varphi}\sin(\theta/2)\\
e^{-i\varphi}\sin(\theta/2)& \cos(\theta/2)\end{array}\right),\quad
U_z(\theta/2)=\left(\begin{array}{cc}e^{i\theta/2}& 0\\ 0& e^{-i\theta/2}\end{array}\right).
\end{equation}
These rotation matrices act on the qubit state $\ket{\psi}$ giving
\begin{eqnarray}\label{rotatedqubit}
\ket{\psi'}&=&U_\mathbf{u}(\theta/2)\ket{\psi}=A_{\theta,\varphi}\ket{\phi}+B_{\theta,\varphi}\ket{\phi+\pi},\nonumber\\
\ket{\psi''}&=&U_z(\theta/2)\ket{\psi}=ae^{i\theta/2}\ket{\phi}+be^{-i\theta/2}\ket{\phi+\pi},
\end{eqnarray}
where
\begin{equation}\label{coeffAandB}
A_{\theta,\varphi}=a\cos(\theta/2)-be^{i\varphi}\sin(\theta/2),\quad
B_{\theta,\varphi}=ae^{-i\varphi}\sin(\theta/2)+b\cos(\theta/2).
\end{equation}

\begin{figure}[pb]
\centerline{\psfig{file=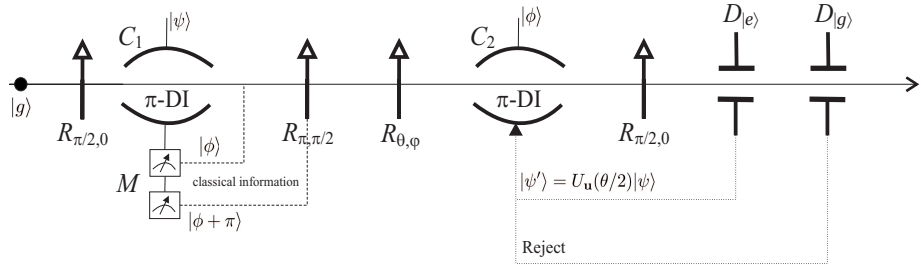}}
\vspace*{8pt}
\caption{\label{fig:rotation}Scheme for realizing a 1-qubit rotation gate about the axis $\mathbf{u}=(-\sin\varphi,\cos\varphi,0)$, $U_\mathbf{u}(\theta/2)$. $C_1$ is initially in the qubit state $\ket{\psi}=a\ket{\phi}+b\ket{\phi+\pi}$ while $C_2$ is prepared in $\ket{\phi}=\ket{0_L}$. $R_{\theta,\varphi}$ represents a Ramsey zone, $M$ the measurement of the cavity state and $D_{\ket{e}}$ ($D_{\ket{g}}$) the excited (ground) atomic state detector. If the final measurement of the atomic state gives $\ket{e}$, the desired rotated qubit $\ket{\psi'}=U_\mathbf{u}(\theta/2)\ket{\psi}$ is then produced in $C_2$. The 1-qubit rotation about the $z$ axis, $\ket{\psi''}=U_z(\theta/2)\ket{\psi}$, is obtained by the same scheme provided that the central Ramsey zone $R_{\theta,\varphi}$ is substituted with two consecutive Ramsey zones $R_{\pi,0}$ and $R_{\pi,\theta/2}$.}
\end{figure}

The rotated qubits of Eq.~(\ref{rotatedqubit}) can be obtained by means of the scheme sketched in Fig.~\ref{fig:rotation}, that requires two cavities $C_1,C_2$: $C_1$ is prepared in the qubit state $\ket{\psi}=a\ket{\phi}+b\ket{\phi+\pi}$, while $C_2$ is in the state $\ket{\phi}$. The protocol for a $U_\mathbf{u}(\theta/2)$ rotation can be described through four steps.

(i) \textit{Copy of the cavity qubit state in an atomic state}. The atom, initially in $\ket{g}$, is prepared in the state $\ket{\chi}=(\ket{g}+\ket{e})/\sqrt{2}$ by the Ramsey zone $R_{\pi/2,0}$ and then enters the cavity $C_1$ for a $\pi$-DI. This interaction gives
\begin{equation}\label{copypsi}
\ket{\chi}\ket{\psi}\stackrel{\pi\textrm{-DI}}{\longrightarrow}\left[\ket{\psi_\textrm{at}}\ket{\phi}
+(\sigma_x\ket{\psi_\textrm{at}})\ket{\phi+\pi}\right]/\sqrt{2},
\end{equation}
where $\ket{\psi_\textrm{at}}=a\ket{g}+b\ket{e}$ is the atomic state copy of the initial cavity qubit state and $\sigma_x$ is the first Pauli matrix. A measurement of the cavity state is then performed. From Eq.~(\ref{copypsi}) it is readily seen that, if the outcome is $\ket{\phi}$, the resulting atomic state is just $\ket{\psi_\textrm{at}}$ and we can pass to the second step. Otherwise, if the outcome is $\ket{\phi+\pi}$, we need to perform a $\sigma_x$ operation to obtain the copy atomic state $\ket{\psi_\textrm{at}}$, achievable up to a unimportant global phase factor by using a Ramsey zone $R_{\pi,\pi/2}$.

(ii) \textit{Atom rotation}. The copy atomic state $\ket{\psi_\textrm{at}}$ is rotated by the Ramsey zone $R_{\theta,\varphi}$ according to Eq.~(\ref{ramseyeq}) to give $\ket{\psi'_\textrm{at}}=A_{\theta,\varphi}\ket{g}+B_{\theta,\varphi}\ket{e}$, where $A_{\theta,\varphi}$, $B_{\theta,\varphi}$ are given in Eq.~(\ref{coeffAandB}).

(iii) \textit{Copy of the atomic state in the cavity qubit state}. The atom is subjected to a $\pi$-DI with the cavity $C_2$ and successively crosses a Ramsey zone $R_{\pi/2,0}$. The total state after these interactions is
\begin{equation}\label{copypsirotated}
\ket{\psi'_\textrm{at}}\ket{\phi}\stackrel{\pi\textrm{-DI}+R_{\pi/2,0}}{\longrightarrow}\left[\ket{\psi'}\ket{e}
+(A_{\theta,\varphi}\ket{\phi}-B_{\theta,\varphi}\ket{\phi+\pi})\ket{g}\right]/\sqrt{2},
\end{equation}
where $\ket{\psi'}=U_\mathbf{u}(\theta/2)\ket{\psi}$ is the desired rotated qubit state given in the first line of Eq.~(\ref{rotatedqubit}).

(iv) \textit{Atomic state detection}. The atomic state is finally measured with the result that, if the outcome is $\ket{e}$, the protocol ends successfully with a probability of $50\%$, as seen from Eq.~(\ref{copypsirotated}).

The rotated qubit state $\ket{\psi''}=U_z(\theta/2)\ket{\psi}$ about the $z$ axis, given in the second line of Eq.~(\ref{rotatedqubit}), is obtained by means of a protocol analogous to that described above, with the only difference that the step (ii) now consists of two consecutive Ramsey zones $R_{\pi,0}$ and $R_{\pi,\theta/2}$. After step (i), the action of these Ramsey zones creates the rotated copy atomic state about the $z$ axis, $\ket{\psi''_\textrm{at}}=ae^{i\theta/2}\ket{g}+be^{-i\theta/2}\ket{e}$. The protocol then ends with the steps (iii) and (iv).

We have thus shown that a 1-qubit rotation can be realized in the CQED framework by 2GBSs and opportune dispersive atom-cavity interactions. Within this context, a Hadamard gate\cite{jeong2002PRA} can be also realized, up to a global phase factor $e^{i\pi/2}$, by three consecutive 1-qubit rotations, namely $H_\textrm{gate}=U_z(\pi/4)U_\mathbf{u}(\pi/4)U_z(\pi/4)$.

Finally, a quantum phase gate (QPG) can be obtained by the protocol described in this section. In particular, the action of the $\pi$-QPG on the qubit state $\ket{\psi}=a\ket{\phi}+b\ket{\phi+\pi}$ is to change of $\pi$ the phase of the logic state $\ket{\phi+\pi}=\ket{1_L}$, giving the output state $\ket{\psi'}=a\ket{\phi}-b\ket{\phi+\pi}$. This $\pi$-QPG action is produced by the steps (i) and (iii) of the above rotation protocol, provided that the final measurement of the atomic state, step (iv), gives $\ket{g}$, as seen from Eq.~(\ref{copypsirotated}) with $A_{\theta,\varphi}=a$ and $B_{\theta,\varphi}=b$.

\section{\label{conclusion}Conclusion}
We have shown that both CNOT and 1-qubit rotation gates can be simply realized in the CQED framework by exploiting orthonormal 2GBSs as qubits stored in high-$Q$ cavities and standard CQED technics as dispersive interactions of Rydberg atoms with the cavities. We have also seen that a Hadamard gate and a quantum phase gate can also be implemented in this context. The CNOT scheme is deterministic (Sec.~\ref{CNOTsection}) while the 1-qubit rotation one is conditional, since it requires the measurement of the final atomic state (Sec.~\ref{singlequbitrotation}). These protocols differ from the ones using coherent states in the optical framework\cite{jeong2002PRA,gilchrist2004JPB,jeong2007}, where both a large photon number and teleportation procedures are required to realize the quantum logic gates.

The practical realization of these quantum gates principally relies on the possibility to produce 2GBSs in a cavity and $\pi$-DI between atom and cavity. In the 1-qubit rotation scheme the measurement of a 2GBS is also needed. The generation and measurement of 2GBSs in a cavity is efficiently achievable by the resonant interaction of two consecutive two-level atoms with a cavity, and it appears to be within the current experimental capabilities\cite{lofranco2006PRA}. The use of $\pi$-DI requires an atom-cavity interaction time $t=\pi\delta/\Omega^2$; this is obtainable with sufficient precision by selecting a suitable atomic velocity $v$, whose typical experimental relative error ($\Delta v/v\approx\Delta t/t\leq10^{-2}$) is small enough not to sensibly affect the schemes\cite{hagley1997PRL}.

In conclusion, the schemes proposed here to realize CNOT and 1-qubit rotation logic gates should be feasible with the established CQED technologies and would provide a set of universal gates for any quantum computation processing.

\section*{Acknowledgments}

G.C. (A.M.) acknowledges partial support by MIUR project II04C0E3F3 (II04C1AF4E) \textit{Collaborazioni Interuniversitarie ed Internazionali tipologia C}.

\vspace*{-6pt}   

\end{document}